\documentclass[a4,11pt]{reportN}
\usepackage[english]{babel}    

\usepackage[utf8]{inputenc}   
\usepackage{indentfirst}
\usepackage{graphics}
\usepackage{subfigure}
\usepackage{graphicx}
\usepackage{epsfig}
\usepackage{psfrag} 
\usepackage[centertags]{amsmath}
\usepackage{graphicx,indentfirst,amsmath,amsfonts,amssymb,amsthm,newlfont}
\usepackage{longtable}
\usepackage[usenames,dvipsnames,svgnames,table]{xcolor}
\usepackage{icomma} 
\usepackage{float} 
\usepackage{verbatim}
\usepackage{multicol}
\usepackage{enumitem}
\usepackage{float}

\usepackage{epstopdf}
\usepackage{psfrag} 

\usepackage[noadjust]{cite}

\begin{document}

\title{Estimating the Largest Lyapunov Exponent Based on Conditional Number}

\author
    { \Large{Pedro Henrique Oliveira Silva}\thanks{pedrolives@hotmail.com.br} \\
  \Large{Vin\'icius da Silva Borges}\thanks{vnisb@hotmail.com} \\
   \Large{Priscila Fernanda da Silva Guedes}\thanks{pri12\_guedes@hotmail.com} \\
    \Large{Igor Carlini Silva}\thanks{igorufsj@yahoo.com.br} \\
     \Large{Erivelton Geraldo Nepomuceno}\thanks{nepomuceno@ufsj.edu.br} \\
     {\small Control and Modelling Group - GCOM, Department of Electrical Engineering,  \\ Federal University of S\~ao Jo\~ao del-Rei, Pra\c{c}a Frei Orlando 170 - Centro, 36307-352 \\
		S\~ao Jo\~ao del-Rei, Minas Gerais, Brasil}\\}

\criartitulo


\begin{abstract}

{\bf Abstract}. The Lyapunov exponent is used to characterize the stability of the dynamic response of the system, and it is often employed to verify if a system is chaotic. Since its discovery in the nineteenth century, various methods have been proposed and developed for its calculation. The present work proposes a method for the calculation of the largest Lyapunov exponent, based on conditional number of the function, which describes the loss of bits in the simulation based on relative rounding error. Four discrete maps are used to show the effectiveness of the proposed method.

\end{abstract}

\section{Introduction}

Today, the chaotic behavior is associated with  dynamics process in several subjects like biology, chemistry, informatics and others. One of the best tools to detect the caos is Lyapunov exponent. The Lyapunov exponents measure the mean divergence or convergence of the near paths along certain directions in space. In chaotic systems, states of two copies of the same system separate exponentially over time, despite very similar initial conditions.

There are several studies regarding the estimation of the Lyapunov exponent \cite{oseledec1968,benettin1980,sano1985,rangarajan1998}. Some of these methods use statistical properties to obtain estimates with high precision  \cite{kantz1994, rosenstein1993practical}, demonstrating the importance of evaluating other concepts to improve understanding and to reevaluate the Lyapunov exponent calculation. These methods to estimate the Lyapunov exponent  have considered  difficult to apply. Recent works have shown new methods, simpler to calculate Lyapunov exponent. For example, Mendes and Nepomuceno \cite{mendes2016very} proposed an algorithm based on the concept of the lower bound error \cite{Nepomuceno2016a}. This method simulates two different interval representations to estimate the Lyapunov exponent, which are the foundations used to calculate the lower bound error representing the error propagation. 

The method proposed by \cite{Nepomuceno2016a} opens a new perspective in the calculation of the Lyapunov exponent, where the limitations  of computational arithmetic is used. Although, this work establishes some relationship between Lyapunov exponent and loss of precision, a clear link has not been demonstrated. To try to fill this gap, this paper presents a procedure to calculate the largest positive Lyapunov exponent (LLE) using the conditional number \cite{ieee}. By doing this, we expect the relationship between the LLE and loss of precision could be more evidenced, as the condition number is precisely defined as the rate of lost digits (or bits) in each iteration. The method was applied to four maps: Logistic, H\'enon, Seno and Lozi, which results have been shown in good agreement with the literature.

\medskip

\section{Preliminary Concepts}

\subsection{Conditional Number}

In solving problems using numerical computation, the conditioning measures how accurate is the solution of the problem using certain floating point precision, independently of the algorithm  and furthermore, represents the loss of significant digits. The evaluation of a real function of a real variable, $y=f(x)$, which in the machine is represented using floating-point arithmetic, is expected to calculate $\hat y= f(\hat x) $. Such that $ \hat x = round(x) $  expresses the rounding error in the representation of the real number $x$, by Overton \cite{ieee}:

\begin{multicols}{2}
     \noindent \begin{equation}
\frac{\hat y -y}{y}\approx k_f(x)\times \frac{|\hat x-x|}{|x|},
	    \label{eq1}
	\end{equation}
	
	\noindent \begin{equation}
k_f = \frac{|x|\times|f^\prime(x)|}{|f(x)|},
	\label{eq2}
	\end{equation}
\end{multicols}

    \noindent where $k_f$ is called a conditional number of $f$ in $x$, where the latter is the initial condition. Basically, the conditional number $k_f$ measures approximately how much the relative rounding error is amplified by the calculation of the function $f$. In Equation \ref{eq1}, the left side of the approximation is a relative measure of how well $y$  approximates $\hat y$, quantified the distance between a reference trajectory denominated as divergence rate, caused by the loss of information. 
    
\begin{equation}
- \log _{2} \left(\dfrac{|\hat y-y|}{|y|}\right) \approx -\log _{2} \left(\dfrac{|\hat x-x|}{|x|}\right) - \log _{2}(k_f(x))
\label{eq3}
\end{equation}

To quantify the divergence, $\log_2$ is used, estimating loss of number of bits at each iteration in the calculation of the function $y$, where IEEE 754 double precision \cite{ieee-754} has initial precision of $53$ bits. Equation \ref{eq3} estimates the number of bits such that $\hat{y} = f (\hat{x})$ agrees with $y = f (x)$, making it possible to calculate the loss of bits, by simply subtracting the precision from the machine with the conditional, $k_f$.






\section{Methodology}

The present work aims to develop a method that is easily implemented for the calculation of the Lyapunov exponent, considering the relative rounding error in the calculation of the functions. The proposed method can be summarized in the following steps:

\begin{enumerate}[label=(\alph*)]
    \item Set the initial conditions and parameters of the chosen function;

    \item Simulate N iterations;

    \item Calculate the conditional number for each value obtained in (b), by Equation \ref{eq1};

    \item Calculate the average of the logarithm with base $2$ of conditional numbers obtained in (c);

    \item The estimation of the Lyapunov exponent is the value obtained in (d).
\end{enumerate}

The magnitude of the Lyapunov exponent quantify the information dynamics  of the observed system. The exponents measure the rate of information created or destroyed in the processing of the system, that is, the loss of bits \cite{kapitaniak1992}. In this way the Lyapunov exponent is expressed in bits of information, where the discrete systems represented have bits/iteration as its units.


The Lyapunov exponent is calculated using $\log$ in the base $2$, which quantifies and defines the loss of bits (information) in each iteration, related to the definition of conditional, which expresses the loss of bits in the simulation by the relative rounding error. The classical Lyapunov exponent estimation procedure was defined by Rosenstein \cite{rosenstein1993practical} and given by Equation \ref{eq4} below


\begin{equation}
d(t)=Ce^{\lambda t}
	\label{eq4}
\end{equation}

    \noindent where $d(t)$ is the average divergence at time $t$, $C$ is a constant that normalizes the initial separation and $\lambda$ is the Lyapunov exponent. The proposed method is a simple strategy to produce a measure related to the average divergence rate in simulations of systems, the conditional number captures the divergence at each iteration related to the loss of precision digits (or bits) and represents the initial separation of the trajectory, similar as shown by Mendes and Nepomuceno \cite{mendes2016very}. 
    
    After that, it is possible to estimate the Lyapunov coefficient, as shown by the classical definition and the concept of conditional number, simplified by the quantification of the divergence rate between a referential trajectory by the rounding error. Table \ref{tab1} presents the values of parameters and the initial conditions found in the literature, used later to compare the performance of the proposed method. The largest Lyapunov exponents (LLE) used in the literature were found in Rosenstein \cite{rosenstein1993practical} (Logistic map), in Wolf \cite{wolf1985determining} (H\'enon map), the Sine map are values found in Mendes and Nepomuceno \cite{mendes2016very} and Lozi map for the works Grassberger and Hata \cite{grassberger1989lyapunov,hata1987} .

\begin{table}[!ht]
\centering
\normalsize
\setlength{\tabcolsep}{10pt} 
\renewcommand{\arraystretch}{1.65}
\caption{Equations of the maps with values of the parameters and initial conditions obtained in the literature.}
\vspace{0.5cm}
\label{tab1}
\begin{tabular}{l l c c c}
\hline
 Systems & Equations & Parameters & $\Delta t (s)$  & Initial conditions \\[1pt] \hline
Logistic & $x_{n+1}=r x_{n}(1-x_{n})$  & $r =4$ & $1$ & $x_{0}=2/3$ \\[1pt] \hline
Sine & $ x_{n+1} = \alpha \sin (x_{n})$  & $ \alpha = 1.2 \pi $ & $1$  & $ x_0=0.1$ \\[1pt] \hline
H\'enon &  $x_{n+1}  = 1 - ax_{n}^2 + y_n $ & $a=1.4$  & $1$ & $x_{0}=0.3$  \\ 
 & $y_{n+1}  = bx_{n} $ & $b=0.3 $  &  & $y_{0}=0.3$ \\[1pt] \hline	 
Lozi &  $x_{n+1}  = 1 - a|x_{n}| + y_n $ & $a=(1+\sqrt{5})/2$  & $1$ & $x_{0}=0.3$  \\[1pt]
 & $y_{n+1}  = bx_{n} $ & $b=-0.3 $  &  & $y_{0}=0.3$ \\[1pt] 	 
 \hline
\end{tabular}
\end{table}


\section{Results}

This section shows the results of applying the method to four discrete time systems, named in Table \ref{tab1}, followed by their respective equations, parameters and initial conditions used in the application of the proposed method. Initially an evaluation was made using the Logistic map shown in Table \ref{tab2}, where the literature has the results of the Lyapunov exponents accompanied by the numbers of iterations necessary for its estimation.

\begin{table}[!ht]
\centering
\normalsize
\setlength{\tabcolsep}{7pt} 
\caption{Comparison of the Lyapunov exponents obtained by Rosenstein \cite{rosenstein1993practical} and the proposed method, using the value $\lambda=0.9999$ (bits/iter.) found by Eckmann \cite{ek1985} to calculate the relative error in the Logistic map.}
\vspace{0.5cm}
\label{tab2}
\renewcommand{\arraystretch}{1.25}
\begin{tabular}{c c r c c c r r}
\cline{1-3} \cline{5-7}
 \multicolumn{3}{c}{Literature} && \multicolumn{3}{c}{Proposed Method}\\
 \cline{1-3} \cline{5-7}
  $\lambda$ (bits/iter.)&  Iterations & Error (\%) & &  $\lambda$ (bits/iter.)  & Iterations & Error (\%) \\[1pt] \cline{1-3} \cline{5-7}
   $0.9507$  & 100 & $-4.9$ && $1.0351$ & $100$ & $3.52$ \\[1pt] 
   $1.0172$  & 200 & $1.7$  && $0.9879$ & $200$ & $ -1.2$ \\[1pt]
   $1.0026$  & 300 & $0.3$  && $0.9969$ & $300$ & $-0.3$ \\[1pt]
   $0.9982$  & 400 & $-0.1$ && $1.0001$ & $400$ & $0.02 $  \\
    \cline{1-3} \cline{5-7}
\end{tabular}
\end{table}

\begin{table}[!ht]
\centering
\normalsize
\setlength{\tabcolsep}{12pt} 
\renewcommand{\arraystretch}{1.15}
\caption{Calculation of the Lyapunov exponent comparing the proposed method with the values obtained in the literature presented by Rosenstein \cite{rosenstein1993practical} (Logistic map), Wolf \cite{wolf1985determining} (H\'enon map), Mendes and Nepomuceno \cite{mendes2016very} (Sine map) and Grassberger and Hata \cite{grassberger1989lyapunov,hata1987} (Lozi map). }
\vspace{0.5cm}
\label{tab3}
\begin{tabular}{l c c c r}
 \hline
 Systems   & Literature  $\lambda$& Calculated $\lambda$ (bits/iter.) & Iterations & Error (\%) \\[5pt]  \hline
Logistic &   $0.9999$ & $1.0351$ & $ 100$ & $3.52$ \\[1pt]
 &  & $0.9879$ & $200$ & $-1.2$  \\[1pt]
 &   & $0.9969$ & $300$ & $-0.3$ \\[1pt]\hline
Sine  &  $1.1550$  & $1.1577$ & $100$ & $ 0.23$ \\[1pt]
 &   & $1.1504$ & $200$ & $-0.40 $  \\[1pt]
 &   & $1.1291$ & $300$ & $ -2.24$ \\[1pt]\hline
H\'enon & $0.6029$  & $0.5963$ & $100$ & $-1.1$ \\[1pt]
 &  & $0.5797$ & $200$ & $-3.86$  \\[1pt]
 &   & $0.5930$ & $300$ & $-1.65$ \\[1pt]\hline
Lozi  & $0.6680$   & $0.6810$ & $100$ & $ 1.94$ \\[1pt]
 &  & $0.6893$ & $200$ & $3.18 $  \\[1pt]
 &   & $0.6896$ & $300$ & $ 3.23$ \\[1pt]
 \hline
\end{tabular}
\end{table}

The number of iterations for an estimate of the Lyapunov exponent of the Logistic map using the proposed method is similar to the numbers obtained by Rosenstein \cite{rosenstein1993practical} (Logistic map), as seen in Table 2. The results were good approximations of those found in the literature, where the same parameters were used for comparison, with showed in Table \ref{tab1}. In the Table \ref{tab3} we showed the calculated $\lambda$ (bit/iter.), loss of bits per iteration, than representation the Largest Lyapunov Exponent (LLE) and the errors of the calculated values in relation to those obtained in the literature, in order to establish performance and legitimacy of the results found by the proposed method.

\begin{figure}[ht!]\centering
\includegraphics[width=0.9\textwidth]{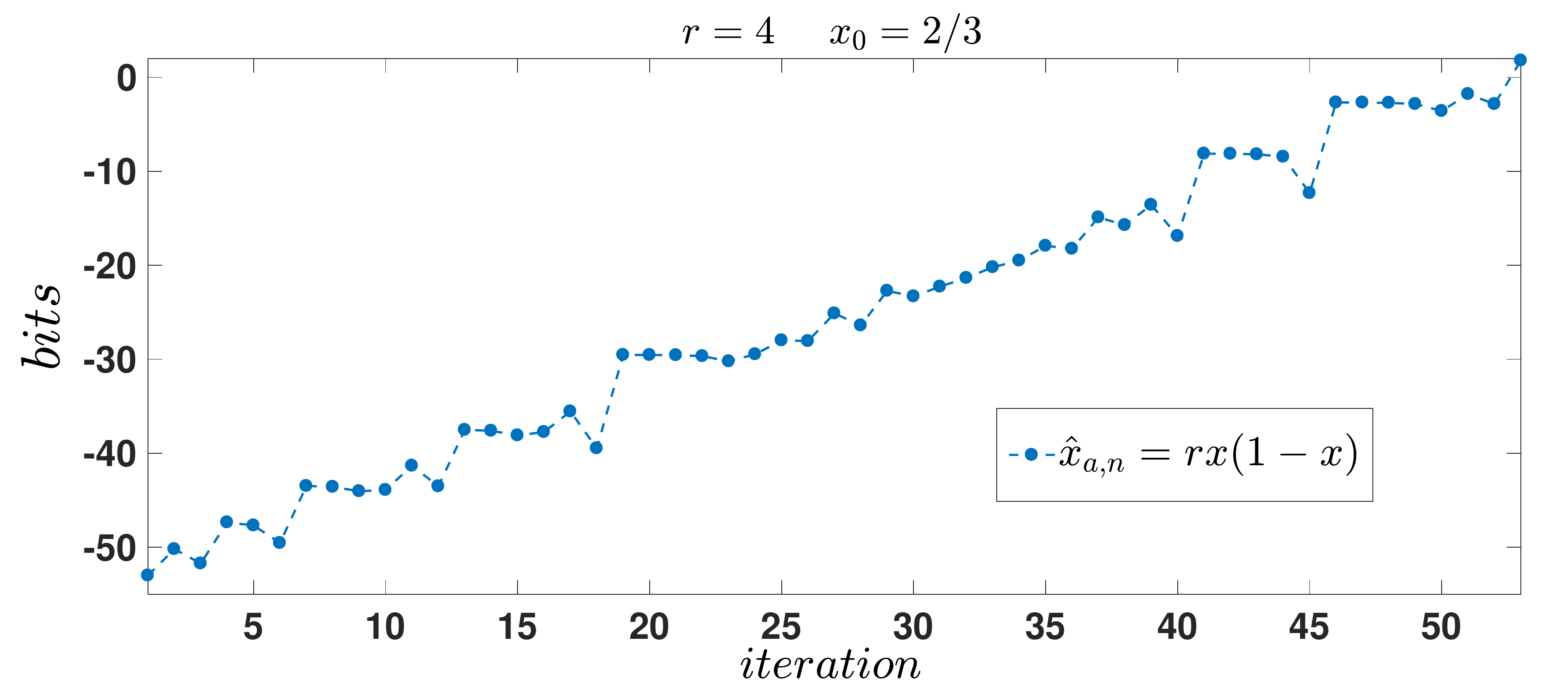}\caption{\label{fig1} Calculation of bit loss at each iteration in the logistic map simulation.}\end{figure}

The proposed method directly relates the rounding error to the Lyapunov exponent concept, representing the loss of bits at each iteration in the calculation of the presented maps. The present work emphasizes the relation of error propagation with the estimation of the exponent using the conditional number as connection. This note is clear in  Figure \ref{fig1}, which shows the loss of accuracy in approximately $53$ iterations, therefore, there is a loss of $1$ bit per iteration in the logistic map  simulation. Note that the loss of precision is easily related to the concept of the Lyapunov exponent, in this example represented by Figure \ref{fig1}, $\lambda$ is $1$ (bit/iteration).


\section{Conclusion}

The proposed method proves the new exposed concept, which establishes the limitation of computational arithmetic relating between the Lyapunov exponent and loss of precision. As demonstrated, the condition number is precisely defined by the digit loss rate per iteration, while containing substantial information about the evolution of the systems. 

The study of chaos detection is directly related to the amplification of the relative rounding error, of the calculation of $f$ in $x$, representing the loss of bits (information) and describes the Lyapunov exponent. The estimates of Lyapunov's exponent using the proposed method were shown to be in a good agreement with the values found in the literature for four well-known chaotic systems. It has also been shown that a original dynamical equation is necessary to estimate the LLE, and that the work is applied in discrete systems. In future work it is intended to be applied to continuous systems.

\section*{Acknowledgments}

The authors are thankful to CAPES, CNPq, FAPEMIG for the support, to UFSJ for the opportunity to develop the project, and to the Control and Modeling Group for the assistance.


\begin{thebibliography}{99}

\bibitem[1]{benettin1980}
G. Benettin, L. Galgani, A. Giorgilli and J. M. Strelcyn, Lyapunov characteristic exponents for smooth dynamical systems and for Hamiltonian systems; A method for computing all of them. Numerical application, Meccanica 15, 21$-$30, (1980), DOI: 10.1007/BF02128236.

\bibitem[2]{ek1985}
J. P. Eckmann and D.  Ruelle, Ergodic theory of chaos and strange attractors, Reviews of modern physics 57(3): 617, (1985), DOI: 10.1103/RevModPhys.57.617. 

\bibitem[3]{grassberger1989lyapunov}
P. Grassberger, On lyapunov and dimension spectra of 2d attractors, with an application to the lozi map, Journal of Physics A: Mathematical and General l22(5): 585, (1989), DOI: 10.1088/0305$-$4470/22/5/020. 

\bibitem[4]{hata1987}
H. Hata, T. Morita, K. Tomita and H.  Mori, Spectra  of  singularities  for  the  lozi and h\'enon maps, Progress of theoretical physics 78(4): 721$-$726, (1987), DOI: 10.1143/PTP.78.721.

\bibitem[5]{kantz1994}
H. Kantz, A robust method to estimate the maximal Lyapunov exponent of a time series, Phys. Lett. A 185, 77$-$87, (1994), DOI: 10.1016/0375$-$9601(94)90991-1. 

\bibitem[6]{kapitaniak1992}
T. Kapitaniak, Chaotic oscillators: theory and applications, World Scientific, vol. 1, (1992).


\bibitem[7]{mendes2016very}
 E. M. Mendes and E.  G. Nepomuceno, A very simple method to calculate the (positive) largest lyapunov exponent using interval  extensions, International  Journal  of Bifurcation and Chaos 26(13): 1650226, (2016), DOI: 10.1142/S0218127416502266.

\bibitem[8]{Nepomuceno2016a}
 E. G. Nepomuceno and  S. A. M. Martins, A lower bound error for free-run simulationof  the  polynomial  narmax, Systems  Science \& Control Engineering, 50$-$58, (2016), DOI: 10.1080/21642583.2016.1163296.
 
\bibitem[9]{oseledec1968}
V. I. Oseledec, The multiplicative ergodic theorem: The Lyapunov characteristic numbers of dynamical systems, Trans. Moscow Math. Soc. 19, 197$-$231, (1968).
 
\bibitem[10]{ieee}
 M. L. Overton, Numerical Computing with IEEE Floating Point Arithmetic, Society for Industrial and Applied Mathematics, Philadelphia, PA, (2001).

\bibitem[11]{rangarajan1998}
G. Rangarajan, S. Habib and R. D. Ryne, Lyapunov exponents without rescaling and reorthogonalization, Phys. Rev. Lett. 80, 3747$-$3750, (1998), DOI: 10.1103/PhysRevLett.80.3747.

\bibitem[12]{rosenstein1993practical}
M.  T. Rosenstein, J.  J. Collins and C. J. De Luca, A practical method for calculating largest lyapunov exponents from small data sets, Physica  D: Nonlinear  Phenomena 65(1$-$2): 117$-$134, (1993), DOI: 10.1016/0167$-$2789(93)90009$-$P.

\bibitem[13]{sano1985}
M. Sano and Y. Sawada, Measurement of the Lyapunov spectrum from a chaotic time series, Phys.
Rev. Lett. 55, 1082$-$1085, (1985), DOI: 10.1103/PhysRevLett.55.1082. 

\bibitem[14]{ieee-754}
M. Standards and C. Society, IEEE Std 754-2008 (Revision of IEEE Std 754-1985), IEEE Standard for Floating-Point Arithmetic, (2008).

\bibitem[15]{wolf1985determining}
A. Wolf, J. B. Swift, H. L. Swinney and J.  A. Vastano, Determining lyapunov  exponents from a time series, Physica D: Nonlinear Phenomena 16(3): 285$-$317, (1985), DOI: 10.1016/0167$-$2789(85)90011$-$9.






\end{thebibliography}
\end{document}